\documentclass[prl,twocolumn,showpacs,floatfix,amsmath,amssymb,superscriptaddress] {revtex4}
\usepackage{amssymb}
\usepackage{latexsym}
\usepackage[letterpaper]{geometry}
\usepackage{latexsym}
\usepackage{graphicx}
\usepackage{times}
\usepackage{amsmath}
\usepackage{dcolumn}
\usepackage{latexsym,amsmath,amssymb,bm,euscript}
\usepackage{epsfig}

\usepackage{amsmath}
\usepackage{amsfonts}

\newcommand{\be}{\begin{equation}}
\newcommand{\ee}{\end{equation}}
\newcommand{\bea}{\begin{eqnarray}}
\newcommand{\eea}{\end{eqnarray}}

\newcommand{\p}{\partial}
\newcommand{\s}{\sigma}

\newcommand{\la}{\langle}
\newcommand{\ra}{\rangle}
\newcommand{\rd}{\mbox{d}}
\newcommand{\ri}{\mbox{i}}
\newcommand{\re}{\mbox{e}}

\begin{document}
\title{
 Composite charge order in the pseudogap region of the cuprates}
\author{ A.M. Tsvelik}
\affiliation{ Department of Condensed Matter Physics and Materials Science, Brookhaven National Laboratory,
  Upton, NY 11973-5000, USA}
  \author{A.V. Chubukov}
  \affiliation{Department of Physics, University of Wisconsin, Madison, WI 53706, USA}
 \date{\today }
 \begin{abstract}
We study the Ginzburg-Landau free energy functional for two coupled $U(1)$  charge order parameters
 describing two non-equivalent charge orders with wave vector ${\bf Q}$ detected in X-ray and STM measurements of underdoped cuprates.
  We do not rely on a mean-field analysis, but rather utilize a field-theoretical technique suitable to study the interplay between vortex physics and
 discrete symmetry breaking in two-dimensional systems with $U(1)$ symmetry.
 Our calculations support the idea that in the clean  systems there are  two transitions: from a high temperature  disordered state into a
  state with a composite
   charge
   order which breaks time-reversal symmetry, but leaves $U(1)$ fields disordered,
    and then into a state with quasi long range order in the $U(1)$ fields.
\end{abstract}


\maketitle
\section{ Introduction.}

 Experimental studies of the cuprates give ample evidence of high complexity of their phase diagram, especially at small doping, where
 superconductivity is preceded by the region of anomalous behavior often called a pseudogap.
 The observations of a  polar Kerr effect~\cite{kerr,he_arpes} and intra-unit-cell magnetic order~\cite{bourges}
  in the  pseudogap region
   indicate that the 
    pseudogap
    is a thermodynamic phase with broken symmetry. 
    This is  at variance, at least partly, with the scenario that the pseudogap is
      just a  precursor to Mott-insulating behavior at and very near half-filling~\cite{mott}.

  Several possible
  symmetry breaking states  
  have been
   proposed 
   for the pseudogap region 
   based on  the analysis of the experimental data.
   In particular, nuclear magnetic resonance (NMR), scanning tunneling microscopy (STM), x-ray, angle resolved photoemission (ARPES), and sound velocity measurements
    in several high-$T_c$ materials ~\cite{ultrasound,mark,davis,ybco,ybco_1,X-ray,X-ray_1,he_arpes,wu},
     were interpreted as
     evidence of sufficiently long-ranged incommensurate charge
      modulations with small momenta $Q_x =(2Q,0)$ and/or $Q_y = (0,2Q)$,
      possibly connecting neighboring hot spots on the Fermi surface~\cite{X-ray}.
       Such modulations necessarily have both charge-density-wave (CDW) and bond-order components~\cite{subir_lar,wang,comin_2}. We will label them as CDW for brevity. Quantum oscillation measurements~\cite{suchitra} and measurements of Hall and Seebeck coefficients~\cite{taillefer} were also interpreted
    as evidence of the feedback effect on fermions from CDW order with small momentum.
       Neutron scattering and other data on La$_{1.875}$Ba$_{0.125}$CuO$_4$ \cite{tranquada} were, on the other hand, interpreted~\cite{kiv_fra} as evidence of 
       long-ranged incommensurate pairing correlations: pair density waves (PDW).
    Both CDW and PDW scenarios were applied~\cite{wang,lee} to fit the ARPES data~\cite{he_arpes} on Pb-Bi2201.
    Other theoretical
     proposals
     for 
      symmetry breaking 
      in the pseudogap phase involve triangular loop currents~\cite{varma} and charge bond order with large momentum near $(\pi,\pi)$~\cite{laughlin}.
\begin{figure}[h]
\begin{center}
\includegraphics[scale=1]{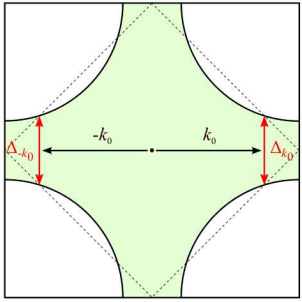}
\caption{
Fermi surface and two CDW order parameters $\Delta_{\bf k} = c^{\dagger}_{k+Q/2} c_{k-Q/2}$ with vertical $Q = (0, Q)$.
 The momenta ${\bf k}$ are located near either ${\bf k}_0$ or $-{\bf k}_0$, and
  $\Delta_{{\bf k}_0}$ and $\Delta_{-{\bf k}_0}$
 are two different $U(1)$ order parameters.}
  \label{fig:FS}
\end{center}
\end{figure}

 A popular theoretical 
  concept behind the
  density-wave order in the cuprates is that this order and superconductivity are
 caused by the same interaction and hence are  intertwined~\cite{kiv_fra_1}, being components of the "super-vector" order parameter (OP).  Along these lines,
 Sachdev and collaborators  argued~\cite{subir_lar,ms,subir_last} that the same  magnetically mediated interaction
 between fermions in hot regions 
  -portions of the Fermi surface (FS) for which antiferromagnetic $(\pi,\pi)$ coincides with $2k_F$,- 
 which  gives rise to $d-$wave superconductivity, also gives rise to  charge bond order (CBO),
   with diagonal momentum ${\bf Q} = (Q, \pm Q)$, and the couplings in the two
  channels are identical if one neglects the curvature of the FS in hot regions.  This idea was further explored by the authors of Ref. \cite{efetov_a},
   who argued that pseudogap behavior may be caused by fluctuations between superconducting and charge order components of the super-vector, whose magnitude
      becomes fixed at $T \sim T^*$, but the direction (superconducting in the presence of FS curvature) is selected only at a much lower $T_c$.

 Three groups recently considered~\cite{wang,subir_a,pepin}
  a charge order with vertical/horizontal ${\bf Q}= (Q,0)$ or $(0,Q)$, as observed in the experiments (see Fig. \ref{fig:FS}).
   The partner of this state (in terms of components of a super-vector) is a
  superconducting PDW order with a non-zero total momentum of a pair~\cite{kiv_fra,lee}.
 The coupling in the CDW/PDW channel is weaker than in the $d$-wave superconducting (SC) channel and, at  first glance, such charge order will be ruined by superconductivity
  and diagonal bond order, which develops at $T_{SC} > T_{CDW}$
 However, the situation is more nuanced for two reasons.
 First, microscopic calculations by two groups show~\cite{wang,sa_ch} that, in the absence of strong superconducting fluctuations or a magnetic field, 
 CDW order appears in the form of a stripe, i.e., with $Q_x$ or $Q_y$ but not with both. Such an order breaks $C_4$ lattice rotational symmetry down to $C_2$.
  At a mean-field level, the $C_4 \to C_2$ symmetry breaking occurs at the onset of long-range charge order, i.e., at $T_{CDW}$. However beyond mean-field,
   the discrete $Z_2$ symmetry associated with the choice of $Q_x$ vs $Q_y$ gets broken at a higher $T_n$ than $T_{CDW}$, much like it  happens for 
    Fe-pnictides~\cite{fe}, where the system breaks $C_4$ lattice rotational symmetry down to $C_2$ before stripe magnetic order sets in.
 If $T_{CDW}$ and $T_{SC}$ are close enough,
    $T_n$ can exceed $T_{SC}$.  Once this happens, the development of a nematic order at $T_n$ produces a negative feedback effect on superconducting $T_c$ and, at the same time, lifts up the temperature of CDW order, $T_{CDW}$, much like the development of a nematic order in Fe-pnictides reduces superconducting $T_c$ and, at the same time, increases the onset temperature for long-range spin order~\cite{fe}.
   
 Second,  for a given $Q_x = (2Q,0)$ or $Q_y = (0,2Q)$, the OP's,  $\Delta_k = c^\dagger_{k+Q}c_{k-Q}$ for CDW and $c^{\dagger}_{k+Q}c^{\dagger}_{k-Q}$ for PDW,
 involve  pairs of hot fermions for which $k$ is located in between two neighboring hot spots, e.g., at
 $2{\bf k} \approx \pm 2{\bf k}_0 = (-2Q,0)$ for ${\bf Q} = (0,Q)$, see Fig. \ref{fig:FS}.
  Because $2{\bf k}_0$ is {\it not} a special symmetry point in the Brillouin zone, the CDW/PDW order parameters
  with ${\bf k} \approx {\bf k}_0$ and ${\bf k} \approx - {\bf k}_0$
  are not equivalent, i.e., even for a given ${\bf Q}$ ($Q_x$ or $Q_y$),  there are two different complex $U(1)$ OP's for CDW~\cite{wang} or PDW~\cite{lee}.
   For the $d$-wave superconducting and charge order with diagonal ${\bf Q}$
   the corresponding ${\bf k}_0 =0$, and there is only one OP.
   The two complex orders for CDW/PDW can 
   potentially
    form another pre-emptive composite order at $T^* > T_{CDW}$, which  
   corresponds to locking the relative phase $\delta \psi$ of the two condensates, while the phase of each condensate still fluctuates freely.
    We will show (see Eq. (\ref{F1}) below) that the  Ginzburg-Landau (GL) functional has minima at $\delta \psi = \pm \pi/2$. The phase locking then
      selects either 
      $+\pi/2$ or $-\pi/2$, giving rise to a non-zero expectation value of $<\Delta_{{\bf k}_0} \Delta^*_{-{\bf k}_0}> = i \Upsilon$ (a four-fermion condensate). Such an order breaks time-reversal symmetry, but preserves parity~\cite{wang}.

In this paper, we use field-theoretical approach, designed specifically to study OP's with $U(1)$ symmetry, and analyze potential pre-emptive composite order,
  which breaks time-reversal symmetry, in the most generic  two-component GL model for CDW/PDW orders with a given $Q_x$ (or $Q_y$). 
  An alternative analysis of such order, based on  Hubbard-Stratonovich transformation to collective variables and subsequent saddle-point analysis, has been presented in~\cite{wang}. The two approaches are complimentary as the  analysis we present here is specific to two-component OP's, while the one  based on  Hubbard-Stratonovich transformation is rigorously justified once one extends the number of components to $M$ and takes  the limit of large $M$. 

   We find that
     a pre-emptive composite order with $<\Delta_{{\bf k}_0} \Delta^*_{-{\bf k}_0}> = i \Upsilon$
     does develop prior to quasi-long-range order in the $U(1)$ fields. For a clean 2D system, we find two  transitions upon lowering $T$ below $T_n$, where
     the system breaks $C_4$ lattice symmetry down to $C_2$ and selects $Q_x$ or $Q_y$.  First  a composite order develops at $T^* \leq T_n$ and time-reversal symmetry gets broken, and then, at a lower $T_{CDW}$, the primary OP's acquire quasi long range order.
  These results are consistent with those obtained using Hubbard-Stratonovich analysis.

  To shorten presentation,  we consider only CDW order. The consideration for PDW order proceeds along the same lines.  A more complicated case
    with  both OP's present is presented in the Appendix D.

\section {The model.}

We assume that $C_4$ lattice symmetry is already broken down to $C_2$ and  consider the
most generic
GL Free energy for two complex order parameters $\Delta_+ = \Delta_{\bf k}$ at ${\bf k} \approx {\bf k}_0$
 and $\Delta_- = \Delta_{\bf k}$ at ${\bf k} \approx {-\bf k}_0$:
\bea
&& {\cal F} = \frac{\rho}{2}\Big(|\p_{\mu}\Delta_{+}|^2 + |\p_{\mu}\Delta_{-}|^2\Big) + \nonumber\\
&& a \Big(|\Delta_{+}|^2 +  |\Delta_{-}|^2\Big) - a'(\Delta^*_{+}\Delta_{-} + \Delta_{+}\Delta^*_{-}) \nonumber\\
&& + b\Big(|\Delta_{+}|^2 + |\Delta_{-}|^2\Big)^2  + c \Big(|\Delta_{+}|^2 - |\Delta_{-}|^2\Big)^2\nonumber\\
&& +d \Big(\Delta^*_{+}\Delta_{-} + \Delta_{+}\Delta^*_{-}\Big)^2
\label{F1}
\eea
where $\mu =x,y$ are spatial components, $a = a(T)$ changes sign at mean-field CDW transition temperature $T_{mf}$,  and we assume that
the prefactors $b,c$ and $d$ are all positive and $b >c$.  The assumption $d >0$ plays a crucial role in our analysis. Indeed,  at $d <0$ only even-in-${\bf k}$ order appears
 via a single transition, and at $d=0$ the system has an extra degeneracy and the pre-emptive order gets destroyed by fluctuations. The GL approach suggests that the order parameter fluctuations are static which is supported by the recent NMR measurements \cite{wu2}.

   The  free energy in this form
has been  derived
   by integrating out fermions either in the spin-fermion model~\cite{wang_2}, or in the model with on-site and nearest-neighbor interaction between fermions~\cite{subir_a,subir_last}.  The $d$ term in (\ref{F1}) shows that the Free energy has minima when the relative phase of $\Delta_{{\bf k}_0}$ and $\Delta_{-{\bf k}_0}$ equals $\pm \pi/2$. A non-zero $a'>0$ implies that 
    the mean-field onset temperature  for $\Delta_+ + \Delta_-$ ( the solution even in ${\bf k}$)
 is larger than that for $\Delta_+ - \Delta_-$ (the solution odd in ${\bf k}$).
 We will demonstrate, that in this situation the only possible spontaneous composite order  is the appearance  of an imaginary  $<\Delta^*_{+}\Delta_{-} - \Delta^*_{-}\Delta_{+}> = i \Upsilon$. This is equivalent to the appearance of a non-zero imaginary $<(\Delta^*_{+} + \Delta^*_{-})*(\Delta_{+} -\Delta_{-})>$.
 Because the second term is odd in ${\bf k}$, and $\Delta_{{\bf k}}$ transforms into $\Delta_{-{\bf k}}$ under time-reversal,  a composite order breaks time-reversal symmetry~\cite{wang}.

   Below the mean field transition temperature 
     the amplitudes of the
   fields $\Delta_+$ and $\Delta_-$
    stabilize and one can replace
 $\Delta_a = |\Delta|z_a$, where $|\Delta|$ is a constant and
 $\sum_{a=+,-}|z_a|^2 =1$,
 and parametrize $z's$ as
\be
 z_+ = \re^{\ri(\phi + \psi)/2}\cos\theta/2, ~~ z_+ = \re^{\ri(\phi - \psi)/2}\sin\theta/2
 \label{param}
\ee
The  Free energy then
can be recast completely in terms of the
 collective variable
$n^a= z^*\s^a z$, where ${\bf n}^2 =1$ as
 (see Appendix A):
\bea
&& F/T = \int \rd^2x \Big\{\frac{1}{2g}[(\p_{\mu}{\bf n})^2 -  (4\pi)^2\vartheta \nabla^{-2}\vartheta] + \nonumber\\
&& \lambda n_z^2 + \kappa n_x + \tau n_x^2\Big\}, \label{F4}
\eea
where $ \vartheta = \frac{1}{4\pi}\epsilon_{\mu\nu}\Big({\bf n}[\p_{\mu}{\bf n}\times\p_{\nu}{\bf n}]\Big)$ is
the topological charge density of the nematic field ${\bf n}$ and
 $g = T/\rho|\Delta|^2, ~~\lambda = c|\Delta|^4/T, ~~\kappa = a'|\Delta|^2/T, ~~\tau = d |\Delta|^4/T$.
The second term in (\ref{F4}) is the Coulomb energy of the topological charges.
 Its presence guarantees that field ${\bf n}$ does not have a finite charge,
 since the energy of such configuration would be infinite. Hence the number of vortices is
   equal to the number of  anti-vortices and there are no topologically nontrivial configurations.

\section {The composite order.}

The composite order is a spontaneous appearance of a non-zero expectation value  $<n_i>$ while $z_+$ and $z_-$ remain strongly fluctuating fields and $<z_{+,-}>=0$.
 The form of $F$ in (\ref{F4}), with positive $\lambda$ and $\tau$, implies
  that only $n_y \propto  i(\Delta^*_{+}\Delta_{-} - \Delta^*_{-}\Delta_{+})$
  can spontaneously acquire a non-zero expectation value.
    Without loss of generality we assume that the anisotropy controlled by $\lambda$-term in (\ref{F4}) is likely to be large so that  the field ${\bf n}$ becomes effectively  two-dimensional. This term also provides a natural scale to the vortex cores $
 R \sim (2\pi/g\lambda)^{1/2}$.
  A large $\lambda$ allows us to
integrate out the fluctuations of the $z$-component of ${\bf n}$ in (\ref{F4}) and derive the effective GL action for the $x-y$ components. This action necessarily includes terms generated by vortices. There are three types of them: (i) $2\pi$-vortices of $\phi$ field combined with a $2\pi$ change of $\theta$, (ii) $2\pi$ vortices of $\psi$ field combined with a $2\pi$ change of $\theta$, and (iii) the joint vortices where $\phi$ and $\psi$ change simultaneously by $2\pi$.  Following  \cite{graphene} we write down the effective free energy functional in the form
\bea
&& {\cal F}/T = \frac{1}{2g}[(\p_x\psi)^2 + (\p_x\phi)^2] + \frac{g}{2}[(\p_x\bar\psi)^2 +\nonumber\\
&& (\p_x\bar\phi)^2] + \ri[\p_x\psi\p_y\bar\psi +\p_x\phi\p_y\bar\phi] + \nonumber\\
&&  \kappa\cos\psi + \tau\cos(2\psi) + \eta_{\psi\theta}\cos(2\pi\bar\psi) +\nonumber\\
&&  \eta_{\phi\theta}\cos(2\pi\bar\phi) + \eta_{\phi\psi}\cos(2\pi\bar\psi)\cos(2\pi\bar\phi) \label{F5}
\eea
where $\bar\psi,\bar\phi$ are  fields dual to $\psi$ and $\phi$ with $\eta_{\alpha\beta}$ being  the corresponding fugacities. The scaling dimensions of the cosine terms are $
d_{\tau} = g/\pi, ~~d_{\phi\theta}=d_{\psi\theta} = \pi/g, ~~d_{\psi\phi} = 2\pi/g$.

 Let us consider the case $\kappa=0$ first. The terms with $\eta_{\phi\theta}, ~\eta_{\psi\theta}$  become irrelevant at $g < \pi/2$, the joint vortices of $\psi$ and $\phi$ become irrelevant at $g < \pi$.  Hence in the region $\pi/2 < g <2\pi$ both $\phi-\theta$ and $\psi-\theta$ vortices and the anisotropy $\tau$-term are relevant and  their competition leads to a transition from a high temperature disordered state to the state with a non-zero  $<n_y>$.

     One can get a general condition for this  transition by equaling the characteristic energy scales generated by the competing  relevant operators:
\be
\tau^{1/(2-d_{\tau})} \sim (\eta_{\psi\theta})^{1/(2-d_{\psi\theta})} \label{condition}
\ee
This condition places the transition temperature $T^*$ in the interval $\pi/2 < g(T^*) < 2\pi$.
  The transition to the state with power-law correlations of $\Delta$ fields on the other hand
  occurs at $T= T_{CDW}$, when  $\phi-\theta$ vortices become irrelevant. This happens when $g (T_{CDW}) =\pi/2$. Because $g(T)$ is an increasing function of $T$,
  the condition
   $g (T^*)>
   g(T_{CDW})$ implies that $T^* > T_{CDW}$.

  If $\tau$ and $\eta_{\psi\theta}$ are comparable, a non-zero $<n_y>$ emerges near $g =\pi$ where it can be studied in detail  as it was done, for instance, in \cite{kuklov}. Near this point  it is convenient to refermionize the part of the effective action containing $\psi,\bar\psi$ fields as  described in Appendix B. The resulting theory is a fermionic model of right- and left moving fermionic Majorana fields $\rho^{(\pm)}_{R,L}$ in 2D with masses $M_{\pm} \sim \tau \pm \eta_{\psi\theta}$ in  Euclidian space governed by the Lagrangian density:
\bea
&& {\cal L}_{\psi} = {\cal L}^{(+)} + {\cal L}^{(-)} + \gamma \rho^{(+)}_R\rho^{(-)}_R\rho^{(+)}_L\rho^{(-)}_L, \nonumber\\
&& {\cal L}^{(a)} = \frac{1}{2}\rho_R^{(a)}(\p_x -\ri\p_y)\rho_R^{(a)} + \nonumber\\
&& \frac{1}{2}\rho_L^{(a)}(\p_x +\ri\p_y)\rho_L^{(a)} + \ri M_a\rho_R^{(a)}\rho_L^{(a)}. \label{ising2}
\eea
 where $a =\pm$ and $\gamma = 4\pi(\pi/g-1)$.
 At high temperatures, the masses $M_{+}$ and $M_{-}$ have different signs.
  The
  transition at $T=T^*$ occurs
   when one of the masses
   passes through zero and changes sign
    (both
      masses
      can never vanish simultaneously). In our case, when $\tau >0$, we found that $M_+$ is positive for all $T$, while $M_-$ changes sign at $T^*$, such that
       both  masses become positive at $T < T^*$.
      The specific heat anomaly at $T$ is only logarithmic  in a clean system $C_v \sim \ln|T-T^*|$ and is further weakened by disorder: $C_v \sim \ln\ln|T-T^*|$ \cite{dotsenko}.

 The CDW OPs and ${\bf n}$  are nonlocal in fermions. However, the theory of free massive Majorana fermions in 2D is equivalent to the 2D Ising model and the corresponding OPs  are expressed in terms of the order and disorder parameter fields of the Ising models $\s$ and $\mu$ \cite{Ising}. The nematic field is
\bea
n_x +\ri n_y \sim \re^{\ri\psi} = (\ri\s_1\s_2 +\mu_1\mu_2). \label{director}
 \eea
At $M_{a} >0$ we have $\la \s\ra \neq 0, \la\mu\ra =0$ and for $M_{a} <0$ we have
$\la \s\ra = 0, \la\mu\ra \neq 0$.
 At $T >T^*$,  when the vortices dominate,
   $M_1$ and $M_2$ have different signs  and both components of $n$ in (\ref{director}) have zero average value.
    At $T< T^*$
     $\s_1$ and $\s_2$
     acquire finite average values,
       and the system develops a
      long range composite
      order in $<n_y>$, which, as we said, breaks time-reversal symmetry.
        The primary CDW OP's however remain disordered as they contain the exponent of $\phi$ field
         whose
          fluctuation remain short ranged for $g > \pi/2$. At $\theta \approx \pi/2$  we have at $T < T^*$
          $
\Delta_+,\Delta_-  = |\Delta|\re^{\ri\phi/2}<\re^{\pm \ri\psi/2}>$ (see Eq. (\ref{param})).
   The CDW transition around $g = \pi/2$ ($T = T_{CDW}$)  is then of the Berezinskii-Kosterlitz-Thouless type and the low temperature phase at $T < T_{CDW}$ is characterized by a quasi-long-range order of the CDW OP's.

 The extension of the results to $\kappa \neq 0$ is straightforward.
 A non-zero $\kappa$ shifts both $T^*$ and $T_{CDW}$ to smaller values and eventually push them into a region of $g$ well below $\pi/2$, where the vortices are irrelevant. In this case, one can analyze the appearance of the composite order semiclassically (see Appendix C).  Still, there are two transitions -- first $<n_y>$ becomes non-zero at $T = T^*$ and then, at $T_{CDW}< T^*$, the correlation length for the principle fields $\Delta_{+,-}$  becomes infinite.

\section{Conclusions}

To conclude, in this paper we studied the GL functional for two coupled $U(1)$  charge order parameters, necessarily present when the order occurs with a vertical/horizontal momentum $Q$ detected in x-ray and STM measurements. 
We considered the temperature range below $T_n$, where the system breaks  $C_4$ lattice rotational symmetry down to $C_2$ and spontaneously selects 
 one of the two potential ordering momenta, either $Q_x = (2Q,0)$ or $Q_y = (0,2Q)$.  
 We used the field-theoretical approach, designed specifically to study order parameters with $U(1)$ symmetry, and analyzed a potential pre-emptive 
  composite order which breaks time-reversal symmetry.
We found that in a  clean 2D system  there are  indeed two transitions: the Ising transition at $T = T^*$, below which the composite order develops 
and time-reversal symmetry gets broken, and a second transition at a lower $T = T_{CDW}$, at which the correlation length for the $U(1)$ order parameters  becomes infinite. 
With application to the hole-doped cuprates, we 
 suggest  that $T^*$ is the onset of the temperature regime where optical measurements show a non-zero polar Kerr effect,  and $T_{CDW}$ is a temperature below which NMR, X-ray, and other measurements detect static CDW order.

A. V. C. is grateful to  K. Efetov, C. Pepin, Y. Wang, S. Kivelson, E. Berg and  S. Lederer for fruitful conversations. He was supported by the DOE grant DE-FG02-ER46900.
A.M.T. was supported by the Center for Emergent Superconductivity, an Energy Frontier Research Center, funded by the Office of Basic Energy Sciences (BES), Division of Materials Science and Engineering, US Department of Energy, through Contract No. DE-AC02-98CH10886.

\section{Appendices}

\subsection{A. Derivation of  free energy (\ref{F4})}

Substituting $\Delta_a = |\Delta|z_a$, where   $|\Delta|$ is a constant and $\sum_{a=s,b}|z_a|^2 =1$, into (\ref{F1}) we obtain the  free energy describing the long wavelength fluctuations as
\bea
&& F/T = \int \rd^2x\Big[\frac{1}{2g}(\p_{\mu}z^*_a\p_{\mu}z_a) + \lambda (z^*\s^zz)^2 \nonumber\\
&& + \kappa (z^*\s^xz) +\tau(z^*\s^xz)^2\Big], \label{F2}
\eea
To obtain (\ref{F4}) we have to integrate over angle $\phi$. The first step is to recast
 (\ref{F2}) as
\bea
&& F/T = \int \rd^2x \Big\{\frac{1}{2g}[(\omega_{\mu}^z)^2 + (\p_{\mu}{\bf n})^2]+ \nonumber\\
&& \lambda n_z^2 + \kappa n_x + \tau n_x^2\Big\}. \label{F3}
\eea
where $\omega_{\mu}^z = \p_{\mu}\phi + \cos\theta\p_{\mu}\psi $.

 Next we integrate over $\phi$ angle. The measure of the path integral is $\rd\Omega = \rd\psi\rd\cos\theta\rd\psi = \rd\phi\rd {\bf n}$, where $n^a = z^*\s^az$.
To integrate over angle $\phi$ with  simple measure $\rd\phi$ we use the following identities;
\bea
 &&\omega_{\mu}^z = \p_{\mu}\alpha + \epsilon_{\mu\nu}\p_{\nu}\chi, ~~ \p^2\chi = 4\pi\vartheta,\label{ident}\\
 &&  \int \rd^2 x (\omega_{\mu}^z)^2 =\int \rd^2x[ (\p_{\mu}\alpha)^2 + (\p_{\mu}\chi)^2]. \nonumber
\eea
The result is Eq. (\ref{F4}) in the main text.

\subsection{B. The fermionic action}

Here we use the standard bosonization formulas. The refermionionized  theory (\ref{F5}) is a fermionic model of right- and left moving fermionic fields $R,L$ in 2D Euclidian space governed by the Lagrangian density:
\bea
&& {\cal L} = {\cal L}_{\phi} + {\cal L}_{\psi}, \nonumber\\
&& {\cal L}_{\psi} = R^+(\p_x -\ri\p_y)R + L^+(\p_x +\ri\p_y)L +\nonumber\\
&&  \ri M_{\tau}(R^+L - L^+R) + M_{\eta}(R^+L^+  + RL) \nonumber\\
&& + 4\pi(\pi/g -1)R^+RL^+L , \label{Ising}\\
&& {\cal L}_{\phi} = \frac{1}{2g}(\p_x\phi)^2 + \frac{g}{2}(\p_x\bar\phi)^2 +\nonumber\\
 && \ri \p_x\phi\p_y\bar\phi  + \eta_{\phi\theta}\cos(2\pi\bar\phi) +... \label{phi}
\eea
where $M_{\tau} \sim \tau, ~~ M_{\eta} \sim \eta_{\psi\theta}$. The dots in (\ref{phi}) stand for the term with $\eta_{\psi\phi}$ which we do not consider as being less relevant. A further simplification occurs when one introduces Majorana fermions
$
R = \frac{1}{\sqrt 2}[\rho_R^{(+)} + \ri\rho^{(-)}_R], ~~ L = \frac{1}{\sqrt 2}[\rho^{(+)}_L + \ri\rho^{(-)}_L]$.
 This leads to Eq. (\ref{ising2}) in the main text.

\subsection{C. The influence of $\kappa$}

In this appendix we discuss
 the role of the $\kappa$ term describing the  difference  in the mean-field transition temperatures 
  for the even- and odd-in-${\bf k}$  components of the CDW OP. The small anisotropy generates the term $\kappa n^x$ in (\ref{F4}) or, equivalently, the term $\kappa\sin\psi$ in (\ref{F5}) and $\kappa \mu_1\mu_2$ in (\ref{ising2}).
This operator does not have a non-zero average in either the ordered or disordered phase of model (\ref{ising2}). Therefore it can play a role only close to the critical point where $|M_-| << M_+$. Since the correlation length $M_+^{-1}$ always remains finite, the $\kappa$ term yields singular correlations only in the second order in $\kappa$ through the fusion:
\bea
&& \mu_1(x)\mu_2(x)\mu_1(0)\mu_2(0) \sim \nonumber\\
&& K_0(M_+x)|x|^{7/8}\rho_R^{(2)}(0)\rho_L^{(2)}(0) +...
\eea
Hence near the critical point the anisotropy generates a contribution to the fermionic Lagrangian $
\ri \kappa^2 M_+^{-9/8}\rho_R^{(2)}\rho_L^{(2)}.$
This corresponds to a shift of $M_-$ meaning a shift of the transition point.

 The increase of $\kappa$ leads to an Ising transition from the nematic phase with a broken time-reversal symmetry to the state where this symmetry is preserved. We will study this transition below $g = \pi/2$ when the vortices are irrelevant. Then the relevant part of  potential in (\ref{F5}) is $
V = \kappa\cos\psi + \tau\cos(2\psi)$.
For $g << 1$ one can analyze this potential semiclassically. At $\kappa =0$ the minimum of the potential is at $\psi = \pi/2$ which corresponds to the nematic director pointing along the $0y$ axis. When $\kappa >0$ increases the minimum moves towards $\psi =\pi$ which corresponds to a reduction of the $y$-component of the director. When $\kappa$ reaches the critical value  $\kappa = 4\tau$ the second derivative at the  minimum  vanishes: $
V = \frac{\tau}{2}(\psi -\pi)^4 + O([\psi-\pi]^6)$,
 which corresponds to the  Ising transition into the state where the director points along the $0x$-axis.

\subsection{D. The interplay between CDW and PDW orders}

As in the main text, we assume that ${\bf Q}$ is either $Q_x$ or $Q_y$.
 The GL action  which includes PDW is
\begin{widetext}
\bea
\frac{\rho}{2}\sum_{a,\alpha}|\p_{\mu}\Delta_{a\alpha}|^2 + a(T- T_{mf})\sum_{a,\alpha}|\Delta_{a\alpha}|^2 + b(\sum_{a,\alpha}|\Delta_{a\alpha}|^2)^2 + c(\Delta^*_{a\alpha}\s^z_{\alpha\beta}\Delta_{a\beta})^2 +d(\Delta_{a\alpha}\s^x_{\alpha\beta}\Delta_{a\beta})^2 \label{F6}
\eea
\end{widetext}
where $a=1,2; \alpha =1,2$. The first index distinguishes between CDW and PDW orders, the second index ($\alpha)$ specifies one of the two components of either CDW or PDW order parameter Well below $T_{mf}$ we can adopt the following parametrization:
\bea
\Delta_{a\alpha} = |\Delta|z_a\left(
\begin{array}{c}
\re^{\ri\psi/2}\\
\re^{-\ri\psi/2}
\end{array}
\right), ~~ \sum_{a=1}^2|z_a|^2 =1. \label{par}
\eea
The symmetry of the resulting long wavelength  model is U(1)$\times$U(1)$\times$SU(2). Substituting (\ref{par}) into (\ref{F6}) we get
\bea
&& {\cal F}/T = \\
&& \frac{1}{8g}(\p_{\mu}\psi)^2 +\lambda\cos(2\psi) +  \frac{1}{2g}(\p_{\mu}z^*_a\p_{\mu}z_a),\nonumber
\eea
($g = 1/\rho|\Delta|^2$, $\lambda = d|\Delta|^4$), which is the sine-Gordon  model plus the SU(2) principal chiral field -the O(4) sigma model. The corresponding OP is a nematic vector
\be
\Delta^*_{a\alpha}\s^{x,y}_{\alpha\beta}\Delta_{a\beta} \sim \exp(\pm i\psi).
\ee
This OP does not contain $z_a$ which always remains disordered since the O(4) sigma model in D=2 always remains in a disordered phase, even in a clean system. If one takes into account vortices in $\psi$, as was done in the main text,  the nematic transition becomes the Ising one. The correlations of the CDW and PDW POs remain short range.

\end{document}